\documentclass[11pt,amstex,aps,amsfonts,prsty,preprint]{article} 
\usepackage[dvips]{graphicx}
\usepackage{amssymb}
\usepackage{amsmath}
\usepackage{a4wide}
\usepackage{citesort}

\begin{document}

\begin{titlepage}

\begin{center}

{\Huge \bf Force-velocity relation and
density profiles for biased diffusion in an adsorbed monolayer}

\vspace{0.3in}

{\Large  O.B{\'e}nichou$^{1,2}$, A.M.Cazabat$^3$, 
J.De Coninck$^{4}$, M.Moreau$^2$ and  G.Oshanin$^2$}

\vspace{0.1in}

{ \sl $^1$ Laboratoire de Physique Th{\'e}orique et Mod{\`e}les Statistiques, \\
Universit{\'e} Paris-Sud, 91405 Orsay Cedex, France
}

\vspace{0.05in}

{ \sl $^2$ Laboratoire de Physique Th{\'e}orique des Liquides, \\
Universit{\'e} Paris 6, 4 Place Jussieu, 75252 Paris, France
}

{ \sl $^3$ Laboratoire de Physique de la Mati{\`e}re Condens{\'e}e, \\
Coll{\`e}ge de France, 11 Place M.Berthelot, 75252 Paris Cedex 05, France
}

{ \sl $^4$ Centre de Recherche en Mod{\'e}lisation Mol{\'e}culaire, \\
Universit{\'e} de Mons-Hainaut, 20 Place du Parc, 7000 Mons, Belgium
}

\begin{abstract}
In this paper, which completes our earlier short publication 
[Phys. Rev. Lett. {\bf 84}, 511 (2000)], 
we study  dynamics of a hard-core tracer particle (TP) performing
a biased random walk in 
an adsorbed monolayer, 
 composed of mobile
hard-core particles undergoing
 continuous exchanges 
with a vapor phase. In terms of
an approximate
approach, based on the decoupling 
of the third-order
correlation functions, 
we obtain the density profiles of the monolayer particles
around the TP and
derive the
force-velocity relation, determining the TP terminal velocity,
$V_{tr}$, as  the function of 
the magnitude of 
 external bias and other
system's parameters. 
Asymptotic forms of the monolayer particles density profiles at large separations
from the TP, and behavior of $V_{tr}$ in the limit of small
external bias are found explicitly.
\end{abstract}

\vspace{0.2in}
\end{center}

\end{titlepage}

\pagebreak

\section{Introduction.}

When a gas  or vapor is brought into contact
with a clean solid surface, some portion of the gas particles   
 becomes reversibly attached to the surface forming an adsorbed layer. 
Following the seminal
work of Langmuir (see, e.g.,  Ref.\cite{surf}), 
equilibrium
 properties of such adsorbed
layers 
have been extensively studied 
and a significant number of
important advancements have been made. 
In particular, 
subsequent analysis
 included more realistic forms of  
intermolecular interactions 
or allowed for a possibility of
multilayer formation.  As a result, 
different phase transformations
have been predicted and 
different forms of
adsorption isotherms 
have been deduced, which are well corroborated by
available experimental data 
(see, e.g.  Refs.\cite{surf,shaw,fowler}).  

Some work has been focused on 
understanding of random motion of individual molecules
in adsorbed layers,  which  constitutes an important factor limiting their
global dynamical behavior. 
For instance, transport processes 
control spreading rates of molecular 
films on solid surfaces \cite{spreading1,spreading}, 
spontaneous or forced dewetting of monolayers 
\cite{aussere,persson,dewetting,dewetting2}
or island formation \cite{islands}.  
Here, some approximate analytical
results have been obtained for
both dynamics of an isolated adatom
on a corrugated surface and collective
diffusion, describing spreading of macroscopic 
density fluctuations in
interacting adsorbates being in contact with 
the vapor phase
\cite{gomer,kreuzer,gortel,wahn}. 
Analysis of tracer diffusion  
in adsorbed layers, which provides a useful information about
intrinsic friction of the adsorbates, pertains mostly 
to two-dimensional models with forbidden 
particles exchanges with the vapor 
(see, e.g. 
Refs.\cite{kehr,nakazato,tahir,beijeren,hilhorst,deconinck}. 
Tracer diffusion in  adsorbed monolayers undergoing 
continuous adsorption and desorption has been essentially less studied. A few available detailed studies
are either devoted to
an analysis of a somewhat artificial
one-dimensional situation 
\cite{benichou}, or,  for two-dimensional adsorbates in contact with the vapor,  
focus on the limit when the tracer particle perform a totally directed random walk \cite{physa}. 
The results for the general situation with arbitrary bias has
been only briefly presented in Ref.\cite{prl}.

In this paper, which completes our earlier short publication  \cite{prl},   
we present a detailed analysis  of 
the properties of a biased tracer diffusion  
in two-dimensional adsorbed monolayers undergoing continuous exchanges
with the vapor. 
The system we consider 
consists of (a)
a solid substrate, which is modeled
in a usual fashion as a regular
square lattice of adsorbtion sites; (b)  
a monolayer of adsorbed, mobile
hard-core particles in contact with a vapor 
and (c) a single hard-core tracer
particle (TP). The  monolayer particles move randomly
along the lattice
by performing symmetric
hopping motion between
the neighboring lattice sites, which process
 is constrained by mutual hard-core interactions, and 
 may desorb from and adsorb onto the lattice from the vapor
with  some prescribed rates dependent on the vapor 
pressure, temperature
 and the interactions with the solid substrate.
In contrast, the
tracer particle is constrained to move 
on the two-dimensional
lattice only, 
(i.e. it can not desorb 
to the vapor), and  is
subject to
a constant external force $E$. Hence, 
the TP performs
a biased random walk,  
constrained by the  hard-core
interactions with the
monolayer particles, 
and always remains within 
the monolayer, probing 
its frictional properties.

In terms of an approximate
approach of Ref.\cite{burlatsky}, based on
the decoupling of the third-order
tracer-particle-particle
 correlation functions into the product of corresponding pairwise 
correlations, 
we determine the density profiles of the monolayer particles, 
as seen from the stationary moving tracer, and 
calculate
analytically
the terminal velocity $V_{tr}$ attained by the tracer particle in the limit $t \to \infty$. 
We show that the monolayer particles distribution  is strongly inhomogeneous:
the local density of the monolayer
 particles in front of the tracer is higher than the 
average, which means that the
monolayer particles
tend to accumulate in front of the driven tracer, 
creating a sort of a "traffic jam", which impedes
 its motion. The condensed, "traffic jam"-like region vanishes 
as an exponential
function of the distance from the tracer. The characteristic
length and the amplitude of the density relaxation function
are calculated explicitly.
On the other hand, past the tracer 
the local density is lower than the average. Here,  we find that depending on
 whether the number of particles in the monolayer 
is explicitly conserved or not, the local density past the tracer
 may tend to the average value at large separations from the tracer 
in a completely different fashion: In the non-conserved case 
the decay of the density is described by an exponential function, while for the conserved particles number case it
shows an
 $\it algebraic$ dependence on the distance, 
revealing in the latter case
especially strong memory effects and strong 
correlations between the particle distribution in the
monolayer and the tracer position. 
Further on, we find
 that the terminal velocity of the tracer particle depends
 explicitly on both the excess density in the "jammed" region in
 front of the tracer, as well as on the density in the depleted region past 
the tracer. We realize that both densities are themselves  dependent on the
 magnitude of the tracer velocity, applied external force,  
as well as on the rate of the adsorption/desorption processes and on the
 rate at which the particles can diffuse away of the tracer, which
results in effective non-linear coupling between $V_{tr}$ and $E$. 
In consequence,    
in the general case (for arbitrary adsorption/desorption 
rates and arbitrary external force), 
$V_{tr}$ can be found only implicitly, 
as the solution of a transcendental 
equation relating $V_{tr}$ to the system parameters. 
This equation 
simplifies considerably   
in the limit of a vanishingly small external bias; 
in this case we obtain  a linear force-velocity relation, 
akin to the so-called Stokes formula, which signifies that in this limit the frictional force exerted on the tracer particle 
by the host medium (the adsorbed monolayer) is viscous.
The corresponding friction coefficient is also determined explicitly.

We finally remark,
that a qualitatively similar physical effect has been predicted recently
for a  different model system involving a charged particle moving 
at a constant speed a small distance above the surface of an incompressible,
 infinitely deep liquid. It has been shown in Refs.\cite{elie1,elie2}, 
that the interactions between the moving particle and the fluid molecules
induce an effective frictional force exerted on the particle, producing
a local distortion of the liquid interface, - a bump, which travels 
together with the particle and increases effectively its mass.  The mass of the bump, 
which is
analogous to the jammed region appearing in our model, depends itself
on the particle's velocity resulting
in a non-linear coupling  between the medium-induced
 frictional force exerted on the particle and its
velocity \cite{elie1,elie2}.

The paper is structured as follows: In Section
 2 we formulate
the model and introduce
basic notations.  In Section 3 we write down the dynamical
equations which govern the time evolution of the monolayer
particles and of the tracer, and outline the decoupling procedure. 
Section 4 is devoted to the
analytical solution of the decoupled discrete-space 
evolution equations in the limit $t \to \infty$ in terms of the generating function approach. 
In Section 5 we analyse 
the asymptotical behavior of the density profiles at large separations in front of and past the tracer particle. 
Next, 
in Section 6,
we derive formal expressions describing the shape of the density profiles in the adsorbed monolayer  
and a force-velocity relation in the general case.
Section 7
 is devoted to the analysis of the general force-velocity 
relation in the limit of a vanishingly small external bias. Here, we derive an analog of the Stokes formula for biased diffusion
in adsorbed monolayers undergoing continuous exchanges with the vapor phase, determine the friction coefficient and estimate,
through the Einstein relation, the tracer diffusion coefficient.
Finally, we conclude in Section 8 with a brief summary
and discussion of our results.

\section{The model}

Consider a
two-dimensional 
square lattice of adsorption sites of spacing $\sigma$, 
which is brought 
in
contact with a reservoir containing identic,
electrically neutral particles (vapor phase) (Fig.1), 
maintained at a constant pressure. 
We suppose that (a) the particles may leave
 the reservoir and adsorb 
onto any vacant lattice site at a fixed rate $f/\tau^*$;  (b) 
the adsorbed particles may move randomly along the lattice by  
hopping at a rate $1/ 4 \tau^*$ to any of four
neighboring sites,
which process is 
constrained by hard-core exclusion preventing multiple occupancy, 
and (c) the adsorbed particles may 
 desorb from the lattice  back to the reservoir
at rate $g/\tau^*$. 
Both $f$ and $g$ are site and environment independent.

To describe the time-dependent occupancy of lattice sites, we introduce 
the
variable $\eta({\bf R})$, which may assume two values:

\begin{equation}
\eta({\bf R}) = \left\{\begin{array}{ll}
1,     \mbox{ if the site ${\bf R}$ is occupied by an adsorbed particle,} \nonumber\\
0,     \mbox{ if the site ${\bf R}$  is empty.}
\end{array}
\right.
\end{equation} 
Evidently, local $\eta({\bf R})$ can change its value due to adsorption,
 desorption and constrained random hopping events, and 
the total number of 
particles in the adsorbed monolayer 
is not explicitly conserved due to adsorption/desorption processes. 
On the other hand, the mean density
 $\rho_s$ of the adsorbate, $\rho_s = <\eta({\bf R})>$, approaches as
$t\to\infty$ a constant value

\begin{equation}
\rho_s=\frac{f}{f+g},
\label{rhostat}
\end{equation} 
which relation represents the customary Langmuir adsorption
isotherm \cite{surf}. We note also that in the analysis of the stationary-state behavior,
 we can always turn to the conserved particles number limit by setting $f$ and $g$ equal to zero and keeping their ratio
fixed, i.e. supposing that $f/g = \rho_s/(1 - \rho_s)$. This limit will correspond to the model of biased tracer diffusion in
a two-dimensional hard-core lattice gas with fixed particles density $\rho_s$, and will allow us to check our analytical
predictions against some already known results \cite{kehr,nakazato,tahir,beijeren,hilhorst,deconinck}.

Further on, at $t = 0$ we introduce at the lattice origin 
an extra hard-core 
particle, whose motion
we would like to follow here and whose 
position at time $t$ we denote as ${\bf R}_{tr}$. This tracer particle - the TP, can be 
thought of as an
external probe designed to
measure the resistance offered by the monolayer
particles to the external perturbance, or, in other words, to measure the intrinsic frictional properties of the adsorbate. 

Now, we stipulate that the TP is different from the adsorbed particles in two aspects:
first, it can not desorb from the lattice and second,  
it experiences an action of
some external driving force, which favors its jumps into a
preferential direction. 
Physically, such a  situation may be realized, 
for instance,
if this only particle is charged and the system is subject to a
uniform electric field {\bf E}.  We suppose here, for simplicity of exposition, that the external force ${\bf E}$
 is oriented in the positive $X$-direction, i.e. ${\bf E} = (E,0)$.

More precisely, we define the TP  dynamics as
follows:
We suppose
that the TP, which occupies the site ${\bf R}_{tr}$ at time $t$,
 waits an exponentially distributed time with mean\footnote{We suppose that in the general case
this mean time  $\tau$ is different from the corresponding time $\tau^*$ 
associated with the monolayer particles dynamics. As a matter of fact,  this should be the case merely because the
TP-substrate interactions may be different from the particle-substrate ones. Varying
 $\tau$ we can mimic different possible situations; in particular, $\tau= 0$ corresponds to the
case when the tracer simply slides along the substrate
 regardless of the surface corrugation. } $\tau$, 
and then attempts to hop onto one of four neighboring sites, ${\bf
R}_{tr} + {\bf e}_{\boldsymbol \nu}$, where ${\bf e}_{\boldsymbol \nu}$ are four unit 
vectors of the square lattice. In what follows we adopt the notation $\nu = \{\pm1,\pm2\}$, where 
${\bf \pm e_1}$ (respectively,  ${\bf \pm e_2}$) will denote 
$\pm X$ (respectively, $\pm Y$) directions.  Next, the jump direction is chosen 
according to the probablity $p_\nu$, which is defined in a usual fashion as
\begin{equation}
p_\nu=\frac{\exp\Big[\frac{\beta}{2}({\bf E \cdot e}_{\boldsymbol
\nu})\Big]}{\sum_{\mu}\exp\Big[\frac{\beta}{2}({\bf E \cdot e}_{\boldsymbol \mu})\Big]},
\label{defp}
\end{equation}    
where $\beta$ is the reciprocal temperature, $({\bf E \cdot e})$ stands for the scalar product, 
the charge of the TP is set equal to unity and  
the sum with the subscript $\mu$ denotes summation over all possible orientations 
of the vector ${\boldsymbol e_\mu}$, that is $\mu = \{\pm1,\pm2\}$.

After the jump direction is chosen, the TP attempts to hop onto the target site. The hop is instantaneously fulfilled 
if 
the target site
is vacant at this moment of time; otherwise, i.e., if the target site
is occupied by any adsorbed particle, the jump is rejected and the
TP remains at its position.

\section{Evolution equations}

Let $\eta\equiv
\{\eta({\bf R})\}$ denote the entire set of the occupation variables, which defines the instantaneous
configuration of the adsorbed particles  at the 
lattice at time moment $t$. Next, let 
$P({\bf R_{tr}},\eta;t)$ denote the joint probability of finding  
at time $t$ the TP at the site ${\bf R_{tr}}$
and all adsorbed particles in the configuration $\eta$.
Then, denoting as $\eta^{{\bf r},\nu}$ a
configuration obtained from $\eta$ by the Kawasaki-type exchange of the occupation
variables of two neighboring 
sites ${\bf r}$ and ${\bf r+e}_{\boldsymbol \nu}$,
and as 
${\hat \eta}^{\bf r}$ - a configuration obtained from 
the original  $\eta$ by
the replacement $\eta({\bf r}) \to 1-\eta({\bf r})$, which corresponds to the
Glauber-type flip of the occupation variable due to 
the adsorption/desorption 
events, we have that 
the time evolution of the configuration
probability $P({\bf R_{tr}},\eta;t)$ obeys the following master equation:

\begin{eqnarray}
&&\partial_tP({\bf R_{tr}},\eta;t)=
\frac{1}{4\tau^*}\sum_{\mu}\;\sum_{{\bf r}\neq{\bf R_{tr}}-{\bf e}_{\boldsymbol \mu},{\bf R_{tr}}}  \; 
\Big\{ P({\bf R_{tr}},\eta^{{\bf r},\mu};t)-P({\bf R_{tr}},\eta;t)\Big\}\nonumber\\
&+&\frac{1}{\tau}\sum_{\mu}p_\mu\Big\{\left(1-\eta({\bf R_{tr}})\right)P({\bf R_{tr}}-{\bf e}_{\boldsymbol \mu},\eta;t)
-\left(1-\eta({\bf R_{tr}}+{\bf e}_{\boldsymbol \mu})\right)P({\bf R_{tr}},\eta;t)\Big\}\nonumber\\
&+&\frac{g}{\tau^*}\sum_{{\bf r}\neq {\bf R_{tr}}} \;\Big\{\left(1-\eta({\bf r})\right)P({\bf R_{tr}},
\hat{\eta}^{{\bf r}};t)-\eta({\bf r})P({\bf R_{tr}},\eta;t)\Big\}\nonumber\\
&+&\frac{f}{\tau^*}\sum_{{\bf r}\neq{\bf R_{tr}}} \;\Big\{\eta({\bf r})P({\bf R_{tr}},\hat{\eta}^{{\bf r}};t)
-\left(1-\eta({\bf r})\right)P({\bf R_{tr}},\eta;t)\Big\},
\label{eqmaitresse}
\end{eqnarray} 
where the subscript ${\bf r} \neq{\bf R_{tr}}$ under the summation symbol signifies 
that summation extends over all lattice sites except
for the site occupied at this time moment by the TP.

Now, the instantaneous 
velocity $V_{tr}(t)$ 
of the TP can be obtained by multiplying both sides of Eq.(\ref{eqmaitresse}) by
$({\bf R_{tr} \cdot e_1})$ and summing over all possible configurations $({\bf R_{tr}},\eta)$.
This results in the following exact equation determining the TP velocity:
\begin{equation}
V_{tr}(t)\equiv\frac{d}{dt} \; \sum_{{\bf R_{tr}},\eta} ({\bf R_{tr} \cdot e_1}) =
\frac{\sigma}{\tau}\Big\{p_1  \Big(1-k({\bf e_1};t)\Big)-p_{-1} \Big(1-k({\bf e_{-1}};t)\Big)\Big\},
\label{vitesse}
\end{equation}
where 
\begin{equation}
k({\boldsymbol \lambda};t)\equiv\sum_{{\bf R_{tr}},\eta}\eta({\bf R_{tr}}+{\boldsymbol \lambda})P({\bf R_{tr}},\eta;t)
\label{defk}
\end{equation}
is the probability of having at time t an adsorbed particle 
at position ${\boldsymbol \lambda}$, defined in the frame of reference moving with the TP. 
Evidently,  $k({\boldsymbol \lambda})$
 can also be interpreted 
as the density
profile in the adsorbed monolayer as seen from the moving TP. 

Equation (\ref{vitesse}) signifies that 
the instantaneous velocity of the TP is dependent on the monolayer particles density in the 
immediate vicinity of the tracer. If the monolayer is perfectly stirred, or, in other words, if 
$k({\boldsymbol \lambda}) = \rho_s$ everywhere, (which implies immediate 
 decoupling of ${\bf R_{tr}}$ and $\eta$),  one would obtain
from Eq.(\ref{vitesse}) a trivial mean-field result
\begin{equation}
V_{tr}^{(0)}=(p_1-p_{-1})(1-\rho_s)\frac{\sigma}{\tau},
\label{vmf}
\end{equation}
which states that the only effect of the medium on the TP dynamics is that its 
jump time $\tau$ gets merely renormalized by a
 factor $(1 - \rho_s)^{-1}$, which represents the inverse 
concentration of voids in the monolayer; note that $(1 - \rho_s)/\tau $ defines simply 
the mean frequency of successful jump events. 
However, the situation appears to be more complicated and, as we proceed to show, 
$k({\boldsymbol \lambda})$ is 
different from the equilibrium value $\rho_s$ everywhere, except for 
 $|\boldsymbol \lambda|\to\infty$. This means that the TP strongly 
perturbs the particles distribution in the monolayer - it is no longer
uniform and some non-trivial stationary density profiles emerge. 
Moreover,
 $k({\boldsymbol \lambda})$ appears to be
dependent 
on the TP velocity which results ultimately  in a 
non-linear coupling between $V_{tr}(t)$ and density profiles.

Now, in order to calculate the instantaneous mean
velocity of the TP we have to determine the mean particles density 
at the neighboring to the TP sites
${\bf R_{tr}}+{\bf e_{\pm1}}$, which requires, in turn, computation of 
the density profile $k({\boldsymbol \lambda})$ for arbitrary $\boldsymbol \lambda$.
The latter can be found
 from the master equation     
(\ref{eqmaitresse}) by multiplying both sides 
by $\eta({\bf R_{tr}})$ and performing the summation over all
configurations $({\bf R_{tr}},\eta)$. In doing so, we
find the following set of equations (see Appendix A for more
details):
\begin{eqnarray}
4\tau^*\partial_tk({\boldsymbol \lambda};t)&=&\sum_{\mu}(\nabla_\mu-\delta_{{\boldsymbol \lambda},{\bf e}_{\boldsymbol 
\mu}}\nabla_{-\mu})k({\boldsymbol \lambda};t)-4(f+g)k({\boldsymbol \lambda};t)+ 4 f\nonumber\\
&+&\frac{4 \tau^*}{\tau}\sum_{{\bf R_{tr}},\eta} \sum_{\mu} p_\mu\left(
1-\eta({\bf R_{tr}+e}_{\boldsymbol \mu})\right)\nabla_\mu\eta({\bf R_{tr}}+{\boldsymbol \lambda})P({\bf R_{tr}},\eta;t),
\label{evolk}
\end{eqnarray}
where  $\nabla_\mu$ denotes the ascending finite difference operator  of the form
\begin{equation}
\nabla_\mu f({\boldsymbol \lambda}) \equiv f({\boldsymbol \lambda}+{\bf e}_{ \mu})-f({\boldsymbol \lambda}),
\label{nabla}
\end{equation}
and
\begin{equation}
\delta_{{\bf r},{\bf r'}} = \left\{\begin{array}{ll}
1,     \mbox{ if the site ${\bf r}={\bf r'},$} \nonumber\\
0,     \mbox{ otherwise.}
\end{array}
\right.
\end{equation} 
The  Kroneker-delta term $\delta_{{\boldsymbol \lambda},{\bf e}_{\boldsymbol \mu}}$ signifies 
that the 
evolution of the pair correlations, Eq.(\ref{evolk}), proceeds differently at large separations and at the
immediate vicinity of the TP, because of 
its asymmetric hopping rules, Eq.(\ref{defp})
(see for more details the points (a) and (b) in the Appendix A).

Note next that the contribution in the second line in Eq.(\ref{evolk}), associated with the TP
 biased diffusion, is non-linear with respect to the occupation numbers such that the
pair correlation function gets effectively coupled to the evolution of the third-order correlations of the
form
\begin{equation}
T({\boldsymbol \lambda},{\bf e}_{\boldsymbol \mu};t)\equiv\sum_{{\bf R_{tr}},\eta}\eta({\bf R_{tr}}+{\boldsymbol
 \lambda})\eta({\bf  R_{tr}}+{\bf e}_{\boldsymbol \mu})P({\bf R_{tr}},\eta;t).
\end{equation}
That is, Eq.(\ref{evolk}) is not closed with respect to the pair correlations but rather represents a
first equation in the infinite hierachy of coupled
equations for higher-order correlation functions. One faces, therefore, the problem of solving an infinite
hierarchy of coupled differential equations and needs to resort to an
approximate closure scheme.

We proceed along the lines suggested in Ref.\cite{prl} and apply
the simplest non-trivial closure approximation, based on the decoupling of the third-order
correlation functions into the product of pair correlations. 
More precisely, we assume that for ${\boldsymbol \lambda}\neq{\bf e}_{\boldsymbol \mu}$ the third-order correlation can be written down in the following
form
\begin{eqnarray}
&&\sum_{{\bf R_{tr}},\eta}\eta({\bf R_{tr}}+{\boldsymbol \lambda})\eta({\bf R_{tr}}+{\bf e}_{\boldsymbol \mu})P({\bf R_{tr}},\eta;t)\nonumber\\
&\approx&\left(\sum_{{\bf R_{tr}},\eta}\eta({\bf R_{tr}}+{\boldsymbol \lambda})P({\bf R_{tr}},\eta;t)\right)\left(\sum_{{\bf R_{tr}},\eta}
\eta({\bf R_{tr}}+{\bf e}_{\boldsymbol \mu})P({\bf R_{tr}},\eta;t)\right)\nonumber\\
&=&k({\boldsymbol \lambda};t)k({\bf e}_{\boldsymbol \mu};t), 
\label{decouplage}
\end{eqnarray}
i.e. the joint 
probability of having at time moment $t$ 
one adsorbed particle in the immediate vicinity of the TP, at position ${\bf e}_{\boldsymbol \mu}$, 
and another particle
at position ${\boldsymbol \lambda}$, is represented as the product of the corresponding pairwise probabilities.
We hasten to remark that the approximate closure of the evolution equations 
in Eq.(\ref{decouplage})  has been already employed for studying related 
models of biased tracer diffusion in hard-core lattices gases and has been
shown to provide quite an
accurate description of both the dynamical and stationary-state behavior.
The decoupling in Eq.(\ref{decouplage})
 has been first introduced in Ref.\cite{burlatsky}
 to determine the properties of a driven 
tracer diffusion in a
one-dimensional hard-core lattice gas with a conserved number of
particles, i.e.  without an exchange of particles with the reservoir.
Extensive numerical simulations performed 
in Ref.\cite{burlatsky} have demonstrated
that such a decoupling provides quite a plausible
approximation for the model under study. 
Moreover,  rigorous
probabilistic analysis of Ref.\cite{olla} has shown 
that for this model the results
 based on the  decoupling scheme in Eq.(\ref{decouplage})
are exact. 
Furthermore, the same closure procedure
has been recently applied  to describe spreading kinetics 
of a hard-core lattice gas from a
reservoir attached to one of the lattice sites \cite{spreading}. Again, a very good
agreement between the analytical predictions and numerical results has been found.
Next, the decoupling in Eq.(\ref{decouplage}) has been used 
in a recent analysis of 
a biased tracer dynamics in
 a one-dimensional model of adsorbed monolayer in contact
with a vapor phase \cite{benichou}, i.e. a one-dimensional version of the model
to be studied here. Also in this case an excellent agreement
has been observed between the analytical 
predictions  and Monte Carlo simulations data \cite{benichou}. 
Besides, as we have already mentioned in Ref.\cite{prl}, the closure of the hierarchy of the evolution equations
  in Eq.(\ref{decouplage}) allows us to reproduce
 in the limit $f,g=0$ and $f/g = const$ (conserved particles number limit)
the results of Refs.\cite{nakazato} and \cite{tahir}, which are known (see e.g. Ref.\cite{kehr}) to provide a very
good approximation for the tracer diffusion coefficient in two-dimensional hard-core
lattice gases with arbitrary particle density.

Using  the approximation in Eq.(\ref{decouplage}), we can rewrite 
Eq.(\ref{evolk}) in the following closed form
\begin{equation}
4\tau^*\partial_tk({\boldsymbol \lambda};t)=\tilde{L}k({\boldsymbol \lambda};t)+4f, 
\label{systemek1}
\end{equation}
which holds for all ${\boldsymbol \lambda}$, except for  ${\boldsymbol \lambda}=\{{\bf 0},\pm{\bf e_1},\pm{\bf e_2}\}$. One the
other hand, for these special sites ${\boldsymbol \lambda} = {\bf e_{\nu}}$ 
with $\nu=\{\pm1,\pm 2\}$ we find
\begin{equation}
4\tau^*\partial_tk({\bf e}_{\boldsymbol \nu};t)=(\tilde{L}+A_\nu)k({\bf e}_{\boldsymbol \nu};t)+4f,  
\label{systemek2}
\end{equation}
where 
$\tilde{L}$ is the operator  
\begin{equation}
\tilde{L}\equiv\sum_\mu A_\mu\nabla_\mu-4(f+g),
\end{equation}
and the coefficients $A_{\mu}$ are defined by
\begin{equation}
A_\mu(t)\equiv1+\frac{4\tau^*}{\tau}p_\mu(1-k({\bf e}_{\boldsymbol \mu};t)).
\label{defA}
\end{equation}

Now, several comments about Eqs.(\ref{systemek1}) and  (\ref{systemek2}) are in order.
First of all, let us note that Eq.(\ref{systemek2}) represents, from the mathematical point of view, 
the boundary conditions for the general evolution equation  (\ref{systemek1}), imposed on the sites in the
immediate vicinity of the TP. As we have noticed already, Eqs.(\ref{systemek1}) and (\ref{systemek2}) have a different
functional form since in the immediate vicinity of the TP  
its asymmetric hopping rules perturb essentially the
monolayer particles dynamics. 

Next, Eqs.(\ref{systemek1}) and  (\ref{systemek2}) possess some
intrinsic symmetries and hence the number of independent parameters can be reduced. Namely, reversing the
field, i.e. 
changing 
$ {\bf E}\to {\bf -E}$,  leads to the mere replacement of
$k({\bf e_1};t)$ by $k({\bf e_{-1}};t)$ but 
does not affect  $k({\bf e}_{\boldsymbol \nu};t)$ with $\nu = \pm 2$, which implies that
  \begin{equation}
k({\bf e_1};t)({\bf -E})=k({\bf e_{-1};t})({\bf E}),\;\;\mbox{and}\;\;k({\bf e}_{ 
\boldsymbol \nu};t)({\bf -E})=k({\bf e}_{\boldsymbol \nu};t)({\bf E}) \;\; \mbox{for}\;\; \nu = \pm 2,
  \label{generalparite}
  \end{equation}
Besides, since  the transition probabilities in Eq.(\ref{defp}) obey
$p_2=p_{-2}$
one evidently has that
\begin{equation}
\label{s1}
k({\bf  e_2};t)=k({\bf e_{-2}};t),
\end{equation} 
and hence, by symmetry, 
\begin{equation}
\label{s2}
A_2(t)=A_{-2}(t)
\end{equation}
which somewhat simplifies Eqs.(\ref{systemek1}) and  (\ref{systemek2}).

Lastly, we note that despite the fact that using the decoupling scheme in
Eq.(\ref{decouplage}) we effectively close the system of equations on the level of the pair
correlations, solution of Eqs.(\ref{systemek1}) and (\ref{systemek2}) still 
poses serious technical 
difficulties: Namely, 
these equations are non-linear with respect to the TP velocity, which enters the gradient term on
the rhs of the evolution equations for the pair correlation, and does depend itself on the values of the
monolayer particles densities in the immediate vicinity of the TP. Below we discuss a solution to this
non-linear problem, focusing on the limit $t \to \infty$.

\section{Stationary solution of the evolution equations}\label{methode}

We turn to the limit $t\to\infty$  and suppose that both the density profiles and stationary velocity of the TP
 have non-trivial stationary values
  \begin{equation}
  k({\boldsymbol \lambda})\equiv\lim_{t\to\infty}k({\boldsymbol \lambda};t), \;\;\;
  V_{tr}\equiv\lim_{t\to\infty}V_{tr}(t),\;\;\;\mbox{and}\;\;\; A_\mu \equiv \lim_{t\to\infty}A_\mu(t)
  \end{equation}
Define next
the local deviations of $k({\boldsymbol \lambda})$ from the unperturbed density as
  \begin{equation}
  h({\boldsymbol \lambda})\equiv k({\boldsymbol \lambda})-\rho_s
  \label{defh}
  \end{equation}
Choosing that $h({\bf 0})=0$, we obtain then the following fundamental
system of equations: 
  \begin{equation}
\tilde{L}h({\boldsymbol \lambda})=0,
  \label{systemeh1}
  \end{equation}
which holds for all ${\boldsymbol \lambda}$ except for ${\boldsymbol \lambda} = \{{\bf 0},{\bf e_{\pm 1}},
{\bf e_{\pm 2}}\}$, 
while for the special sites adjacent to the TP, i.e. for ${\boldsymbol \lambda} = \{{\bf 0},{\bf e}_{\pm 1},
{\bf e}_{\pm 2}\}$, one has
  \begin{equation}
 (\tilde{L}+A_\nu)h({\bf e}_{\boldsymbol \nu})+\rho_s(A_\nu-A_{-\nu})=0,
  \label{systemeh2}
  \end{equation}
Equations (\ref{systemeh1}) and (\ref{systemeh2})
determine the deviation from the unperturbed density $\rho_s$ 
in the stationary state.  Note also that in virtue of the symmetry relations in Eqs.(\ref{s1}) and (\ref{s2}), 
$h({\bf e_{2}}) =h({\bf e_{-2}})$
and $A_2 = A_{-2} $.

Now, our general approach to solution of coupled non-linear 
Eqs.(\ref{vitesse}),(\ref{systemeh1}) and (\ref{systemeh2})
 is as follows:  We first derive a general solution of these equations
supposing that $V_{tr}$ is a given parameter, or, in other words, 
assuming that the coefficients $A_\nu$ 
entering Eqs.(\ref{systemeh1}) and (\ref{systemeh2}) are known.
In doing so, we obtain $h({\boldsymbol \lambda})$ in the parametrized form
\begin{equation}
h({\boldsymbol \lambda})=h({\boldsymbol \lambda}; A_{\pm 1},A_{2}).
\label{hA}
\end{equation}
Then, substituting into Eq.(\ref{hA}) particular values ${\boldsymbol \lambda}=\{{\bf e}_{\pm 1},{\bf e}_{\pm 2}\}$ 
and making use of the definition of $A_\mu$ in 
Eq.(\ref{defA}), we find a system of three linear equations with three
 unknown coefficients of the form
  \begin{equation}
A_\nu=1+\frac{4 \tau^*}{\tau}p_\nu\Big(1-\rho_s-h({\bf e}_{\boldsymbol \nu}; 
A_{\pm 1},A_{ 2})\Big),
  \end{equation}
where  $\nu=\{\pm1,2\}$, 
which will allow us to define all $A_\nu$ (and hence, all
$h(\bf e_{\boldsymbol \nu}))$. 
 Finally, substituting the results into Eq.(\ref{vitesse}), which can be written down in
terms of $A_\nu$ as
  \begin{equation}
  V_{tr}=\frac{\sigma}{4\tau^*}(A_1-A_{-1}),
  \label{vitimp}
  \end{equation}
we will arrive at a closed-form equation determining implicitly 
the stationary velocity.  

\subsection{Generating function for the stationary particle density profiles in the monolayer.}

Equations (\ref{systemeh1}) and (\ref{systemeh2}) can be most conveniently solved by introducing the 
generating function for the density profiles of
the form 
\begin{equation}
H(w_1,w_2)\equiv\sum_{n_1=-\infty}^{+\infty}\sum_{n_2=-\infty}^{+\infty}h_{n_1,n_2}w_1^{n_1}w_2^{n_2},
\label{defH}
\end{equation}
where $h_{n_1,n_2}\equiv h(n_1{\bf e_1}+n_2{\bf e_2})$.   
Multiplying both sides of Eqs.(\ref{systemeh1}) and (\ref{systemeh2}) by $w_1^{n_1}w_2^{n_2}$,
 and performing summations over $n_1$ and $n_2$ 
we find that $H(w_1,w_2)$ is  given explicitly by
\begin{eqnarray}
H(w_1,w_2)&=& - \; K(w_1,w_2)\;\Big\{A_1w_{1}^{-1}+A_{-1}w_1+A_2(w_2+w_2^{-1})-\alpha\Big\}^{-1}, 
\label{adevelopper}
\end{eqnarray}
where
\begin{equation}
\alpha\equiv\sum_\nu A_\nu+4(f+g),
\end{equation}
and 
\begin{equation}
K(w_1,w_2)\equiv \sum_\nu
A_\nu(w_{|\nu|}^{\nu/|\nu|}-1)h({\bf e}_{\boldsymbol
\nu})+\rho_s(A_1-A_{-1})(w_{1}-w_1^{-1})
\label{defK}. 
\end{equation}
Equations (\ref{adevelopper}) to (\ref{defK}) determine the generation function for the density profiles exactly, and
the latter can be explicitly obtained via standard inversion formulae.

\subsection{Integral characteristic of the density profiles}

As we have already remarked,  the presence of the driven TP
 induces
 an inhomogeneous density distribution
in the monolayer. One can thus pose a natural question whether 
equilibrium between adsorption and desorption processes gets shifted due to such a perturbancy, 
i.e. whether 
the equilibrium density in the monolayer is different from that given by Eq.(\ref{rhostat}). The answer is trivially "no"
 in the case when the particles number is
explicitly conserved, but in the general case 
with arbitrary $f$ and $g$ this is not at all evident: similarly to the behavior in one-dimensional system \cite{benichou},
one expects that also in two-dimensions the density  profiles are asymmetric as seen from the stationary moving 
TP and are characterized by a
condensed, "traffic-jam"-like region in front of and a depleted region past the TP. One anticipates then that the
desorption events are 
favored in front of the TP, while  the adsorption events are evidently suppressed by the excess density.
On the other hand, past the TP desorption is diminished due to 
the particles depletion while adsorption
may proceed more readily due to the same reason.
 It is thus not at all clear $\it a \; priori$ whether these two effects can compensate
each other exactly, in view of a possible asymmetry of the density profiles, as it happens in the one-dimensional model \cite{benichou}.

For this purpose, we study the behavior of  the integral
deviation $\Omega$ of the density from the equilibrium value $\rho_{s}$, i.e.
\begin{equation}
\Omega\equiv\sum_{n_1=-\infty}^{+\infty}\sum_{n_2=-\infty}^{+\infty}h_{n_1,n_2},
\end{equation}
which can be computed straightforwardly from
 Eqs.(\ref{adevelopper}) and  (\ref{defK}) by setting both $w_1$ and $w_2$ equal to unity.  Noticing that
$K(w_1=1,w_2=1)=0$, and that
$A_1+A_{-1}+2A_2-\alpha=-4(f+g)$, i.e. is strictly negative as soon as 
adsorption/desorption processes are present, we obtain then
that $\Omega$ is stricly equal to 0.  This implies, in turn,  that the 
perturbancy of the density distribution in the monolayer created by the driven TP
does not shift the global balance between the adsorption and desorption events. 
 An analogous  result has been obtained for the
one-dimensional problem in Ref.\cite{benichou}. 

\subsection{Inversion of the generating function with respect to the "symmetric" coordinate $w_2$.}

Inversion of the generating function with respect to $w_2$ can be readily performed by expanding $H(w_1,w_2)$ into the series
 in powers of
$w_2$. To do this, we first rewrite $H(w_1,w_2)$ as 
\begin{eqnarray}
H(w_1,w_2)&=&\alpha^{-1} K(w_1,w_2)\sum_{i=0}^{+\infty}\alpha^{-i}\Big(A_1w_{1}^{-1}+A_{-1}w_1+A_2(w_2+w_2^{-1})\Big)^i.
\label{dev}
\end{eqnarray}
Then, the sum on the right-hand-side of Eq.(\ref{dev}) can be further expanded as
\begin{eqnarray}
&&\sum_{i=0}^{+\infty}\alpha^{-i}\Big(A_1w_{1}^{-1}+A_{-1}w_1+A_2(w_2+w_2^{-1})\Big)^i\nonumber\\
&=&\sum_{n_2=-\infty}^{+\infty}\sum_{i=0}^{+\infty}\sum_{j=0}^{+\infty}\alpha^{-(i+|n_2|+2j)}\binom{i+|n_2|+2j}{|n_2|+2j}\binom{|n_2|+2j}{j}(A_1w_{1}^{-1}+A_{-1}w_1)^iA_2^{|n_2|+2j}w_2^{n_2},\nonumber\\
\label{calcul1}
\end{eqnarray}
where $\binom{n}{p}$ stands for the binomial coefficients. Next, gathering the terms with the same power of $w_2$ on the two sides of
Eq. (\ref{dev}) and using Eq. (\ref{calcul1}), we have that
\begin{eqnarray}
\sum_{n_1=-\infty}^{+\infty}h_{n_1,n_2}w_1^{n_1}
&=&\alpha^{-1}\Big\{-\sum_\nu A_\nu h({\bf e}_{\boldsymbol \nu})+\Big(A_1h({\bf e_1})+\rho_s(A_1-A_{-1})\Big)w_1\nonumber\\
&+&\Big(A_{-1}h({\bf e_{-1}})-\rho_s(A_1-A_{-1})\Big)w_1^{-1}\Big\}F^{(2)}(w_1,|n_2|)\nonumber\\
&+&\alpha^{-1}A_2h({\bf e_2})\Big(F^{(2)}(w_1,|n_2-1|)+F^{(2)}(w_1,|n_2+1|)\Big),
\label{sert}
\end{eqnarray}
where
\begin{equation}
F^{(2)}(w_1,n_2)\equiv\sum_{i=0}^{+\infty}\sum_{j=0}^{+\infty}\alpha^{-(i+n_2+2j)}\binom{i+n_2+2j}{n_2+2j}\binom{n_2+2j}{j}A_2^{n_2+2j}(A_1w_{1}^{-1}+A_{-1}w_1)^i.
\label{repserF2}
\end{equation}

As the next step, we resum the series entering $F^{(2)}(w_1,n_2)$. 
Substituting to  Eq.(\ref{repserF2}) explicit expressions for  the binomial coefficients 
and for the  Gamma function, we have 
 \begin{eqnarray}
F^{(2)}(w_1,n_2)
&=&\left(\frac{A_2}{\alpha}\right)^{n_2}\sum_{j=0}^{+\infty}\Big\{\left(\frac{A_2}{\alpha}\right)^{2j}\frac{1}{\Gamma(j+1)\Gamma(n_2+j+1)}\nonumber\\
&\times&\int_0^{+\infty}e^{-t}t^{n_2+2j}\sum_{i=0}^{+\infty}\left(\frac{A_1w_{1}^{-1}
+A_{-1}w_1}{\alpha}\right)^i\frac{t^i}{\Gamma(i+1)}{\rm d}t\Big\},
\end{eqnarray}  
which yields
\begin{eqnarray}
F^{(2)}(w_1,n_2)
&=&\left(\frac{A_2}{\alpha}\right)^{n_2}\left(1-\frac{A_1w_{1}^{-1}+A_{-1}w_1}{\alpha}\right)^{-(n_2+1)}\nonumber\\
&\times&\sum_{j=0}^{+\infty}\left(\frac{A_2}{\alpha-(A_1w_{1}^{-1}+A_{-1}w_1)}\right)^{2j}\binom{n_2+2j}{j}.
\end{eqnarray}
Next, using the identity
\begin{equation}
\sum_{j=0}^{+\infty}x^{2j}\binom{n_2+2j}{j}=2^{n_2}(1-4x^2)^{-1/2}\left(1+\sqrt{1-4x^2}\right)^{-n_2},
\end{equation} 
we obtain 
\begin{equation}
F^{(2)}(w_1,n_2)=\frac{\alpha y}{A_2\sqrt{1-4y^2}}\left(\frac{2y}{1+\sqrt{1-4y^2}}\right)^{n_2},
\label{y}
\end{equation}
where $y=A_2/(\alpha-(A_1w_{1}^{-1}+A_{-1}w_1))$. Eqs.(\ref{sert}) to (\ref{y}) define the inverted with respect to $w_2$ generating function.

\subsection{Inversion of the generating function with respect to the
"asymmetric" coordinate $w_1$.}

Inversion of $H(w_1,w_2)$ with respect to $w_1$ 
can be performed along exactly the same lines as we did it in the previous subsection
for the variable
 $w_2$. That is, by expanding $F^{(2)}(w_1,n_2)$ in Eq.(\ref{y}) into a series
in powers of $w_1$:
\begin{eqnarray}
&&F^{(2)}(w_1,n_2)=\left(A_{sign(-n_1)}\right)^{|n_1|}\sum_{n_1=-\infty}^{+\infty}\sum_{k=0}^{+\infty}\sum_{j=0}^{+\infty}
\Big\{\alpha^{-(|n_1|+n_2+2j+2k)} \times \nonumber\\ 
&\times&\binom{|n_1|+n_2+2j+2k}{n_2+2j}\binom{n_2+2j}{j}\binom{|n_1|+2k}{k}
 A_2^{n_2+2j}(A_1A_{-1})^k w_1^{n_1}\Big\}.
\end{eqnarray}
and then identifying the coefficients in this expansion with the analogous coefficients in the
expansion in powers of the variable $w_1$ of  
\begin{eqnarray}
h_{n_1,n_2}=
\alpha^{-1}\Big\{\sum_\nu A_\nu h({\bf e}_{\boldsymbol \nu})\nabla_{-\nu}F_{n_1,n_2}-\rho_s(A_1-A_{-1})(\nabla_1-\nabla_{-1})F_{n_1,n_2}\Big\}.
\label{2dsolh}
\end{eqnarray}
with
\begin{eqnarray}
F_{n_1,n_2}&\equiv&\left(A_{sign(-n_1)}\right)^{|n_1|}\sum_{k=0}^{+\infty}\sum_{j=0}^{+\infty}\Big\{\alpha^{-(|n_1|+|n_2|+2j+2k)}
\times \nonumber\\
&&\binom{|n_1|+|n_2|+2j+2k}{|n_2|+2j}\binom{|n_2|+2j}{j}
\binom{|n_1|+2k}{k}A_2^{|n_2|+2j}(A_1A_{-1})^{k}\Big\}
\label{repser}
\end{eqnarray}
Then, using an integral identity
\begin{eqnarray}
&&\binom{|n_1|+|n_2|+2j+2k}{|n_2|+2j}\binom{|n_2|+2j}{j}\binom{n_1+2k}{k}\nonumber\\
&=&\int_0^{\infty}e^{-t}t^{|n_1|+|n_2|+2k+2j+1}\frac{{\rm d}t}{\Gamma(k+1)\Gamma(|n_1|+k+1)\Gamma(j+1)\Gamma(|n_2|+j+1)} \nonumber\\
\end{eqnarray}
we cast $F_{n_1,n_2}$ into the form
\begin{eqnarray}
F_{n_1,n_2}&=&
\left(\frac{A_{-1}}{A_1}\right)^{n_1/2}\int_0^{\infty}e^{-t}{\rm I}_{n_1}\left(2\alpha^{-1}\sqrt{A_1A_{-1}}t\right){\rm I}_{n_2}\left(2\alpha^{-1}A_2t\right){\rm d}t,
\label{repint}
\end{eqnarray}
where ${\rm I}_n(z)$ stands for the  modified Bessel, defined as
\begin{equation}
\label{i}
{\rm I}_n(z) = \frac{1}{2 \pi} \int^{\pi}_{-\pi} d\theta \; \cos(n \theta) \exp(z \cos(\theta))
\end{equation}
It is worth-while to mention now that $F_{n_1,n_2}$ 
has an interesting physical interpretation. To illustrate it, 
we rewrite the 
integral involved in Eq.(\ref{repint}) 
using the definition in Eq.(\ref{i}) as 
\begin{equation}
\int_0^{\infty}e^{-t}{\rm I}_{n_1}\left(2\alpha^{-1}\sqrt{A_1A_{-1}}t\right){\rm I}_{n_2}\left(2\alpha^{-1}A_2t\right){\rm d}t=\frac{1}{(2\pi)^2}\int_{-\pi}^{\pi}\int_{-\pi}^{\pi}\frac{e^{-(k_1n_1+k_2n_2)}}{1-\xi\lambda({\bf k})}{\rm d}k_1{\rm d}k_2,
\label{2repint}
\end{equation} 
where 
\begin{equation}
\xi=2\alpha^{-1}(\sqrt{A_1A_{-1}}+A_2),\;\;\mbox{ and }\;\;\lambda({\bf k})=\frac{\sqrt{A_1A_{-1}}\cos(k_1)+A_2\cos(k_2)}{\sqrt{A_1A_{-1}}+A_2}.
\label{parammarche}
\end{equation}  
One recognizes then (cf. ref. \cite{Hughes}) that the rhs of
Eq. (\ref{2repint}) is the generating function
\begin{equation}
P(n_1,n_2;\xi)\equiv\sum_{j=0}^{+\infty}P_j(n_1,n_2)\xi^j,  \;\;\;\xi<1, 
\label{defP}
\end{equation}
of $P_j(n_1,n_2)$ -  the probability that a particle 
 performing some special type of random walk on the sites of a
two-dimensional square lattice and starting at the origin  arrives exactly on the $j$-th step to the
site with the lattice vector $n_1{\bf e_1}+n_2{\bf e_2}$. This special random walk is characterized by
$\lambda({\bf k})$ - the structure factor
(cf. ref. \cite{Hughes}), which defines a biased random walk with the following elementary jump
probabilities 
\begin{equation}
\frac{\sqrt{A_1A_{-1}}}{2\left(\sqrt{A_1A_{-1}}+A_2\right)},\;\;\mbox{
in the  directions  }\;\;\pm{\bf e_1}\;\;\mbox{of the lattice}, 
\end{equation}     
\begin{equation}
\frac{A_2}{2\left(\sqrt{A_1A_{-1}}+A_2\right)},\;\;\mbox{ in the
directions  }\;\;\pm{\bf e_2}\;\;\mbox{of the lattice}. 
\end{equation}
Hence, $F_{n_1,n_2}$ can be interpreted as
\begin{equation}
F_{n_1,n_2}=\left(\frac{A_{-1}}{A_1}\right)^{n_1/2}P(n_1,n_2;\xi),
\label{rwsimple}
\end{equation}
and thus can be thought of as the generating function of a two-dimensional biased random walk. 

\section{Asymptotical behavior of the density profiles at large distances from the tracer particle.}

Equations (\ref{sert}) to (\ref{y}) allow us to deduce an asymptotical behavior of the
 density profiles in front of and past the stationary moving TP. To do this, 
note first that these asymptotic forms of $h_{n,0}$ with $n\to\pm\infty$ can be most straightforwardly obtained 
from the generating function of $h_{n,0}$:
\begin{equation}
N(z)\equiv\sum_{n=-\infty}^{+\infty}h_{n,0}z^n,
\label{2ddefN}
\end{equation}
which we proceed now to calculate. Using Eqs.(\ref{sert}) and (\ref{y}), we find that
\begin{eqnarray}
N(z)&=&\frac{z(z-1)\Big(A_1h({\bf e_1})+\rho_s(A_1-A_{-1})\Big)+(1-z)\Big(A_{-1}h({\bf
e_{-1}})-\rho_s(A_1-A_{-1})\Big)}{A_{-1}\sqrt{(z-z_1)(z-z_2)(z-z_3)(z-z_4)}}\nonumber\\
&&\;+h({\bf e_2})\left(\sqrt{\frac{(z-z_2)(z-z_3)}{(z-z_1)(z-z_4)}}-1\right),
\label{expressionN}
\end{eqnarray} 
where the roots $z_i$ are defined as
\begin{equation}
\label{2ddefz1}
z_1=\frac{1}{A_{-1}}\left\{\frac{\alpha}{2}+A_2-\sqrt{\left(\frac{\alpha}{2}+A_2\right)^2-A_1A_{-1}}\right\},
\end{equation}
\begin{equation}
z_2=\frac{1}{A_{-1}}\left\{\frac{\alpha}{2}-A_2-\sqrt{\left(\frac{\alpha}{2}-A_2\right)^2-A_1A_{-1}}\right\},
\label{z2}
\end{equation}
\begin{equation}
z_3=\frac{1}{A_{-1}}\left\{\frac{\alpha}{2}-A_2+\sqrt{\left(\frac{\alpha}{2}-A_2\right)^2-A_1A_{-1}}\right\},
\end{equation}
\begin{equation}
\label{2ddefz4}
z_4=\frac{1}{A_{-1}}\left\{\frac{\alpha}{2}+A_2+\sqrt{\left(\frac{\alpha}{2}+A_2\right)^2-A_1A_{-1}}\right\},
\end{equation}
which obey
\begin{equation}
0<z_1\leq z_2\leq1<z_3< z_4.
\end{equation}
In consequence,  $N(z)$ is a holomorphic function in the annular region of inner radius $z_2$ 
and the outer radius $z_3$. As explained in Appendix B, the behavior of  $h_{n,0}$ when $n\to+\infty$
 (resp. $n\to-\infty$) is controled by $z_3$ (resp. $z_2^{-1}$).

\subsection{Asymptotics of density profiles at large distances in front of the tracer particle.}

The asymptotical behavior of the density profiles in front of the TP is supported
by the behavior of the generating function $N(z)$ in the vincinity of
$z_3$ (see Apendix B for more details). We find then that  
for $n\to+\infty$ the density obeys
\begin{equation}
h_{n,0}\sim\frac{z_3(z_3-1)\Big(A_1h_{\bf e_1}+\rho_s(A_1-A_{-1})\Big)+(1-z_3)\Big(A_{-1}h_{\bf e_{-1}}-\rho_s(A_1-A_{-1})\Big)}{A_{-1}\sqrt{(z_3-z_1)(z_3-z_2)(z_4-z_3)}}\frac{1}{\sqrt{\pi n}}\frac{1}{z_3^n},
\end{equation}
which means that in front of the TP the deviation $h_{n,0}$  decays exponentially with the distance,
\begin{equation}
h_{n,0}\sim K_+\frac{\exp\Big(-n/ \lambda_+\Big)}{n^{1/2}},
\label{2ddevh}
\end{equation}
where the decay amplitude obeys
\begin{eqnarray}
K_+&=&\frac{(z_3-1)\Big(A_1h({\bf e_1})+\rho_s(A_1-A_{-1})\Big)+(1/z_3-1)\Big(A_{-1}h({\bf e_{-1}})
-\rho_s(A_1-A_{-1})\Big)}{\sqrt{8\pi A_2}}\nonumber\\
&&\;\times\left\{\left(\frac{\alpha/2-A_2}{A_{-1}}\right)^2-\frac{A_1}{A_{-1}}\right\}^{-1/4},
\label{2damplidevant}
\end{eqnarray}
while the characteristic decay length is given by
\begin{equation}
\lambda_+=1/ \ln(z_3).  
\label{2dlongueurdevant}
\end{equation}
Note that here $\lambda_+$ is finite for any values of the system parameters.

\subsection{Asymptotics of density profiles at large distances past the tracer particle.}

We first note that  one of the  roots of the generating function, 
namely $z_2$, gets equal to unity when both $f$ and $g$ are strictly 
equal to zero, which results in the exact cancellation of the
multiplier $\sqrt{1 - z}$ both in the nominator and the denominator in the
first term on the rhs of
Eq.(\ref{expressionN}).
 This signifies that 
in the limit when exchanges with the reservoir are forbidden, 
the singular behavior of the generating
function at
the vicinity of $z_2$ is essentially different compared to the case when both $f$ and $g$ are greater than zero. 
Consequently, one has to consider separately 
the behavior in the case of non-conserved particles number, when
exchanges with the
reservoir persist, and the case when both $f$ and $g$ are equal to
 zero while their ratio is kept fixed.

\paragraph{A. Non-conserved particles number.}

In case when both $f$ and $g$ have non-zero values, we find  (cf. Appendix B) that when $n\to+\infty$ the density deviation 
follows
\begin{equation}
h_{-n,0}\sim K_-\frac{\exp\Big(-n/ \lambda_-\Big)}{n^{1/2}},
\label{2dderh}
\end{equation}
where the amplitude is given by
\begin{eqnarray}
K_-&=&\frac{(z_2-1)\Big(A_1h_{\bf e_1}+\rho_s(A_1-A_{-1})\Big)+(1/z_2-1)\Big(A_{-1}h_{\bf e_{-1}}-\rho_s(A_1-A_{-1})\Big)}{\sqrt{8\pi A_2}}\nonumber\\
&&\;\times\left\{\left(\frac{\alpha/2-A_2}{A_{1}}\right)^2-\frac{A_{-1}}{A_{1}}\right\}^{-1/4},
\label{2dnonconservampliderriere}
\end{eqnarray}
and the characteristic decay length obeys
\begin{equation}
\lambda_-=-1/ \ln(z_2).  
\label{2dlongueurderriere}
\end{equation}
Consequently, in the case when particles exchanges with the reservoir are
permitted, the density deviation 
from the equilibrium value
$\rho_s$ decays exponentially with the distance from the TP. 
Note that the decay
lenghts satisfy the inequality $\lambda_->\lambda_+$, which means that the
correlations between the tracer and particles of the monolayer are
always stronger past than in front of the TP. Note also that $\lambda_-$ diverges when both $f$ and $g$ tend to zero, 
which signals, as we have already
remarked, that the decay of
$h_{n,0}$ may proceed differently in this case, compared to the exponential dependence in Eq.(\ref{2dderh}). 

\paragraph {B. Conserved particles number.}

Suppose now that both $f$ and $g$ are equal
 to zero,  while their ratio is
fixed and obeys $f/g = \rho_s/(1-\rho_s)$.
 As we have already remarked, this
 situation corresponds to the
customary model of a two-dimensional hard-core lattice gas without exchanges with a 
reservoir. For this situation, we find (see Appendix B for more details), that 
when $n\to+\infty$ the deviation of the particle density from the equilibrium value $\rho_s$ obeys
\begin{equation}
h_{-n,0}=-\frac{K_-'}{n^{3/2}}\left(1+\frac{3}{8n}+{\mathcal O}\Big(\frac{1}{n^2}\Big)\right),
\label{algebrique}
\end{equation} 
where the decay amplitude $K_-'$ is given by
\begin{eqnarray}
K_-'&=&\frac{1}{4\sqrt{\pi}}\left\{\left(A_1h({\bf e_1})-A_{-1}h({\bf e_{-1}})+2\rho_s(A_1
-A_{-1})\right)\sqrt{\frac{1}{A_2(A_1-A_{-1})}}\right.\nonumber\\
&&\left.\;+h({\bf e_2})\sqrt{\frac{A_1-A_{-1}}{A_2}}\right\}
\label{2dconservampliderriere}
\end{eqnarray}
Remarkably enough, in this case the correlations between the TP position
 and the particles distribution
vanish  {\it algebraically} slow with the distance! This implies, in turn,
 that in the conserved particles
number case, the mixing of the monolayer is not efficient enough to prevent the appearence of the quasi-long-range order
and the medium "remembers"
 the passage of the TP
on a long time and space scale, which signifies very 
strong memory effects. We note also that the algebraic
 decay of correlations 
in this model has been predicted earlier in Ref.\cite{deconinck}. However, the
 decay exponent has been
erroneously suggested to be equal to $1/2$, as opposed to the value
 $3/2$ given by Eq.(\ref{algebrique}).  
As well, the amplitude $K_-'$ happens to have a different sign, 
compared to that obtained in Ref.\cite{deconinck},
which invalidates the conclusion 
that the overall relaxation 
to the equilibrium value $\rho_s$ might show a non-monotoneous behavior 
as a function of the distance past the TP.

\section{Formal expression for the density profiles
and general force-velocity relation.}

The Eqs.(\ref{2dsolh}) and (\ref{repint}) display $h_{n_1,n_2}$
as a function of the coefficients $A_\nu$ that remain to be determined.
As a matter of fact, these coefficients depend  themselves
on the local densities in the immediate vicintiy of the tracer, i.e. on $h({\bf e_\nu})$. 
This implies that we have to determine them  from Eqs.(\ref{2dsolh}) and (\ref{repint}) in a self-consistent way.

Setting in  Eq.(\ref{2dsolh}) ${\boldsymbol \lambda} = {\bf e}_{\boldsymbol \nu}$, where $\nu=\{\pm1,2\}$,
results in the
following system of equations
\begin{equation}
\tilde{C}\tilde{h}=\rho_s(A_1-A_{-1})\tilde{F},
\end{equation}
where
\begin{equation}
\tilde{h}=
\begin{pmatrix}
h({\bf e_1})\\
h({\bf e_{-1}})\\
h({\bf e_2})
\end{pmatrix},
\;\;\tilde{F}=
\begin{pmatrix}
(\nabla_1-\nabla_{-1})F_{\bf e_1}\\
(\nabla_1-\nabla_{-1})F_{\bf e_{-1}}\\
(\nabla_1-\nabla_{-1})F_{\bf e_2}
\end{pmatrix},
\end{equation}
and
\begin{equation}
\tilde{C}=
\begin{pmatrix}
A_1\nabla_{-1}F_{\bf e_1}-\alpha & A_{-1}\nabla_1F_{\bf e_1} & A_{2}\nabla_{-2}F_{\bf e_1}\\
A_1\nabla_{-1}F_{\bf e_{-1}} & A_{-1}\nabla_1F_{\bf e_{-1}}-\alpha & A_{2}\nabla_{-2}F_{\bf e_{-1}}\\
A_1\nabla_{-1}F_{\bf e_2} & A_{-1}\nabla_1F_{\bf e_2} &  A_{2}\nabla_{-2}F_{\bf e_2}-\alpha
\end{pmatrix}.
\end{equation}
The solutions of this  $3\times3$ linear system  are
\begin{equation}
\forall\nu=\{\pm1,2\},\;\;\;\;h({\bf e}_{\boldsymbol \nu})=\rho_s(A_1-A_{-1})\frac{\det\tilde{C}_\nu}{\det\tilde{C}},
\label{formelle}
\end{equation}
where $\tilde{C_\nu}$ stands for the matrix obtained from  $\tilde{C}$
by replacing the $\nu$-th column  by the column-vector
$\tilde{F}$. The substitution of expression in Eq.(\ref{formelle}) into
the definition of the coefficients $A_\nu$, (\ref{defA}), results in the
following system of three equations 
\begin{equation}
\forall\nu=\{\pm1,2\},\;\;\;\;A_\nu=1+\frac{4\tau^*}{\tau}p_\nu\left\{1-\rho_s-\rho_s(A_1-A_{-1})\frac{\det\tilde{C}_\nu}{\det\tilde{C}}\right\},
\label{2dimplicite}
\end{equation}  
which implicitly determines all unknown coefficients $A_\nu$ and hence, 
the local deviations $h({\bf e}_{\boldsymbol \nu})$, defined
as
\begin{equation}
h({\bf e}_{\boldsymbol \nu})=(1-\rho_s)+\frac{\tau}{4\tau^*p_\nu}(1-A_\nu)
\label{ppasnul}.
\end{equation}  
Finally, for every lattice point, except for the origin occupied by the TP, the density profiles
are given by Eq.(\ref{2dsolh}), where  $F_{n_1,n_2}$ is defined by
Eq. (\ref{repint}).

Next, substituting Eq.(\ref{formelle}) to Eqs.(\ref{vitesse}) and (\ref{defh}), we find that the TP terminal
velocity obeys
\begin{equation}
V_{tr}=\frac{\sigma}{\tau}\Big\{(p_1-p_{-1})(1-\rho_s)-\rho_s(A_1-A_{-1})\frac{p_1\det\tilde{C}_1-p_{-1}\det\tilde{C}_{-1}}{\det\tilde{C}}\Big\},
\end{equation}
which can be rewritten, taking into account  Eq.(\ref{vitimp}), as  
\begin{equation}
V_{tr}=\frac{\sigma}{\tau}(p_1-p_{-1})(1-\rho_s)\Big\{1+\rho_s\frac{4\tau^*}{\tau}\frac{p_1\det\tilde{C}_1-p_{-1}\det\tilde{C}_{-1}}{\det\tilde{C}}\Big\}^{-1}.
\label{forcevitesse}
\end{equation}
This last equation represents the 
desired general force-velocity relation for the system under study, which is valid for arbitrary
magnitude of external bias and arbitrary values of 
other system's parameters.  In the general form as it is, Eq.(\ref{forcevitesse}) is apparently
not very useful, but may describe physically interesting behavior in two particular 
limits - the case of infinitely strong bias,
when the TP performs totally directed random walk, which is appropriate to motion of the AFM tip, 
and the case when the bias
is vanishingly small. While the former case has been studied in detail in Ref.\cite{physa}, the results for the
latter case have been only briefly outlined in Ref.\cite{prl}. In the next Section we present detailed derivation of the
limiting form of the force-velocity relation in case of a vanishingly small external force.

\section{Vanishingly small external bias.}

We turn now to the limit $\beta E \ll 1$, in which case the problem simplifies considerably and allows to obtain explicit results
for the local densities in the immediate vicinity of the TP and consequently, the TP terminal velocity and
diffusivity.

\subsection{Local particle density in the vicinity of the tracer in the limit $\beta E \ll 1$.}

Setting ${\boldsymbol \lambda} = {\bf e_1}$ and ${\boldsymbol \lambda} = {\bf e_{-1}}$
in Eq.(\ref{2dsolh}), we obtain equations obeyed by 
 $h({\bf e_{\pm1}})$:
\begin{equation}
\alpha h({\bf e_1})=\sum_\nu A_\nu h({\bf e_{\boldsymbol \nu}})
\nabla_{-\nu}F_{\bf e_1}-\rho_s(A_1-A_{-1})(\nabla_1-\nabla_{-1})F_{\bf e_1},
\label{fre1}
\end{equation}
and
\begin{equation}
\alpha h({\bf e_{-1}})=\sum_\nu A_\nu h({\bf e_{\boldsymbol \nu}})
\nabla_{-\nu}F_{\bf e_{-1}}-\rho_s(A_1-A_{-1})(\nabla_1-\nabla_{-1})F_{\bf e_{-1}}.
\label{frem1}
\end{equation}
Consequently, the discontinuity in the density profile is given by:
\begin{eqnarray}
\delta h = \Big(h({\bf e_1})-h({\bf e_{-1}})\Big) &=& \alpha^{-1} \sum_\nu A_\nu h({\bf e_{\boldsymbol \nu}})
\nabla_{-\nu}(F_{\bf e_1}-F_{\bf e_{-1}})-\nonumber\\
&-&\rho_s(A_1-A_{-1})(\nabla_1-\nabla_{-1})(F_{\bf e_1}-F_{\bf e_{-1}})
\label{fricdiff}
\end{eqnarray}
Our aim is now to compute the leading order contribution to  $\delta h$ in the limit $\beta E \ll 1$.
To this purpose, we first 
expand $p_\nu$, $h({\bf e_{\pm 1}})$ and $A_\nu$ in the Taylor series in powers
 of $E$ retaining only linear with the field terms:
\begin{equation}
p_\nu=\frac{1}{4}\Big(1+\frac{\beta\sigma}{2}{\bf (E \cdot e_{\boldsymbol \nu})}+{\mathcal O}(E^2)\Big),
\end{equation}
\begin{equation}
h_{\bf e_1}\propto E\;\;\;\mbox{and}\;\;\;h_{\bf e_{-1}}=-h_{\bf e_1}+{\mathcal O}(E^2),
\label{paire}
\end{equation}
\begin{equation}
\label{paire2}
h_{\bf e_{\pm2}}={\mathcal O}(E^2),
\end{equation}
and consequently, 
\begin{equation}
A_\nu=1+\frac{\tau^*}{\tau}(1-\rho_s)+\frac{\tau^*}{\tau}\Big((1-\rho_s)\frac{\beta\sigma}{2}{\bf 
(E \cdot e_{\boldsymbol \nu})}-h({\bf e}_{\boldsymbol \nu})\Big)+{\mathcal O}(E^2), 
\end{equation}
where we have made use of the
symmetry relations given by Eq.  (\ref{generalparite}). Using these
equations, we can expand the terms in the right-hand-side of Eq.(\ref{fricdiff}) as:
\begin{eqnarray}
\sum_\nu A_\nu h({\bf e}_{\boldsymbol \nu})\nabla_{-\nu}(F_{\bf e_1}&-&F_{\bf e_{-1}})=
\sum_\nu A_\nu h({\bf e}_{\boldsymbol \nu})\nabla_{-\nu}(F_{\bf e_1}(E=0)-F_{\bf e_{-1}}(E=0))+{\mathcal O}(E^2)\nonumber\\
&=&h_{\bf e_1}(A_1\nabla_{-1}-A_{-1}\nabla_1)(F_{\bf e_1}(E=0)-F_{\bf e_{-1}}(E=0))
+{\mathcal O}(E^2)\nonumber\\
&=&\frac{2h({\bf e_1})}{{\cal L}(2A_0/ \alpha_0)}\left(1+\frac{\tau^*}{\tau}(1-\rho_s)\right)+{\mathcal O}(E^2),
\label{frdev1}
\end{eqnarray}
where
\begin{equation}
{\cal L}(x)\equiv\left\{\int_0^\infty e^{-t}\Big(({\rm I}_0(xt)-{\rm I}_2(xt)){\rm I}_0(xt){\rm d}t\right\}^{-1},
\end{equation}
\begin{equation}
A_0\equiv A_\nu(E=0)=1+\frac{\tau^*}{\tau}(1-\rho_s),
\end{equation}
and $\alpha_0\equiv\alpha(E=0)$. Next, we have that 
\begin{eqnarray}
\rho_s(A_1-A_{-1})(\nabla_1-\nabla_{-1})(F_{\bf e_1}&-&
F_{\bf e_{-1}})=\nonumber\\
&-&\frac{2\rho_s}{{\cal L}(2A_0/\alpha_0)}\frac{\tau^*}{\tau}\left((1-\rho_s)\beta\sigma E-2h({\bf e_1})\right)+{\mathcal O}(E^2)
\label{frdev2}
\end{eqnarray}
Substituting Eqs.(\ref{frdev1}) and (\ref{frdev2}) into Eq.(\ref{fricdiff}),  we obtain
\begin{eqnarray}
h({\bf e_1})&=&\frac{h({\bf e_1})}{\alpha_0 {\cal L}(2A_0/
\alpha_0)}\left(1+\frac{\tau^*}{\tau}(1-\rho_s)\right)+\nonumber\\
&+&\frac{\rho_s}{\alpha_0{\cal L}(2A_0/ \alpha_0)}\frac{\tau^*}{\tau}\left((1-\rho_s)
\beta\sigma E-2h({\bf e_1})\right)+{\mathcal O}(E^2),
\end{eqnarray}
which leads to the desired result for the deviation of the local
 density just in front of the TP from the equilibrium value
$\rho_s$. The leading order contribution to $h({\bf e_1})$ with respect to the field is thus given explicitly by 
\begin{equation}
h({\bf e_1})=\rho_s(1-\rho_s)\beta\sigma E\frac{\tau^*}{\tau}\left\{\alpha_0{\cal L}(2A_0/ \alpha_0)-1+(3\rho_s-1)\frac{\tau^*}{\tau}\right\}^{-1}+{\mathcal O}(E^2).
\label{fricequivh}
\end{equation}
Hence, the discontinuity $\delta h$ 
in the monolayer particles density in the immediate vicinity of the TP
equals, by virtue of Eq.(\ref{paire}), twice the expression on the right-hand-side of Eq.(\ref{fricequivh}).

\subsection{Friction coefficient and the Stokes formula.}
Expanding the TP velocity $V_{tr}$ in the Taylor
 series in powers of $E$ and again, retaining only linear with $E$ terms, we
get
\begin{equation}
V_{tr}=\frac{\beta\sigma^2}{4\tau}(1-\rho_s)E-\frac{\sigma}{4\tau}\Big(h({\bf e_1})-h({\bf e_{-1}})\Big)+{\mathcal O}(E^2)
\end{equation}
Next, making use of Eqs.(\ref{paire}) and  (\ref{fricequivh}), we
arrive at the following explicit result
\begin{equation}
\label{v}
V_{tr}\sim\frac{\beta\sigma^2}{4\tau}(1
-\rho_s)E\left\{1-\frac{2\rho_s\tau^*}{\tau}\frac{1}{\alpha_0{\cal L}(2A_0/ \alpha_0)-1+(3\rho_s-1)\tau^*/\tau}\right\},
\end{equation} 
which signifies that in the limit of a vanishingly small external bias the frictional 
force exerted on the TP by the monolayer particles
is viscous.
Note also that Eq.(\ref{v}) 
is quite similar to the well-known  Stokes formula, i.e. $V_{tr}\sim E/\zeta$, and in our case the friction coefficient 
$\zeta$ is given explicitly by
\begin{equation}
\zeta=\frac{4\tau}{\beta\sigma^2(1-\rho_s)}\left\{1+\frac{\tau^*}{\tau}\frac{2\rho_s}{\alpha_0{\cal L}(2A_0/ \alpha_0)-A_0}\right\}.
\end{equation} 
Note now that the friction coefficient $\zeta$ is the sum of two
contributions. 
The first one,
\begin{equation}
\zeta_{cm}\equiv\frac{4\tau}{\beta\sigma^2(1-\rho_s)},
\end{equation}
is a typical mean-field-type result and corresponds to a perfectly homogeneous monolayer (see discussion following Eq.(6)).
The second one, 
\begin{equation}
\zeta_{coop}=\frac{4\tau^*}{\beta\sigma^2(1-\rho_s)}\frac{2\rho_s}{\alpha_0{\cal L}_2(2A_0/ \alpha_0)-A_0},
\end{equation} 
has, however, a more complicated origin.  Namely, it is associated with the 
cooperative behavior emerging in the monolayer -
dehomogenization of the particle distribution in
the adsorbed monolayer due to the presence of a driven impurity (the TP)
 and formation of stationary density profiles,
whose characteristic properties depend on the velocity $V_{tr}$.

\subsection{Diffusion coefficient and Einstein relation}

Lastly, assuming {\it a priori} that the Einstein relation holds for the system
under study, we estimate the TP diffusion coefficient $D_{tr}$ as
\begin{equation}
D_{tr}=\beta^{-1}\zeta^{-1}=\frac{\sigma^2}{4\tau}(1-\rho_s)\left\{1-\frac{2\rho_s\tau^*}{\tau}\frac{1}{\alpha_0{\cal L}(2A_0/ \alpha_0)-1+(3\rho_s-1)\tau^*/\tau}\right\}.
\label{2dautodiffgen}
\end{equation}
It seems now interesting to compare our general result in Eq.(\ref{2dautodiffgen}) against the classical result of Nakazato and
Kitahara \cite{nakazato}, which describes TP
 diffusion coefficient in a two-dimensional lattice gas with conserved
particles number. 
Setting $f$ and $g$ equal to zero, while assuming that their ratio 
has a  fixed value, $f/g=\rho_s/(1-\rho_s)$, we have then 
that 
\begin{equation}
\label{d}
D_{tr}\to\hat{D}_{tr}=\frac{\sigma^2}{4\tau}(1-\rho_s)\left\{1-\frac{2\rho_s\tau^*}{\tau}\frac{1}{4A_0{\cal L}(1/2)-1+(3\rho_s-1)\tau^*/\tau}\right\}.
\end{equation}
Noticing next that
\begin{equation}
\frac{1}{{\cal L}(1/2)}=\lim_{\xi\to1^-}\Big(P(0,0;\xi)-P(2,0;\xi)\Big),
\end{equation}
where $P(n_1,n_2;\xi)$ has been defined by Eq.(\ref{defP}), and using
the fact that \cite{mccrea}
\begin{equation}
\lim_{\xi\to1^-}\Big(P(0,0;\xi)-P(2,0;\xi)\Big)=4-\frac{8}{\pi},
\end{equation}
we find that the right-hand-side of Eq.(\ref{d}) attains the form
\begin{equation}
\label{comparNaka}
\hat{D}_{tr}=\frac{\sigma^2}{4\tau}(1-\rho_s)\left\{1-
\frac{2\rho_s\tau^*}{\tau}\frac{1-2/\pi}{1+(1-\rho_s)\tau^*/\tau-(1-2/\pi)(1+(1-3\rho_s)\tau^*/\tau)}\right\},
\end{equation}
which expression
 is exactly the same as the one obtained earlier in 
Refs.\cite{nakazato} and \cite{tahir} within the framework of different, compared to our approach,
 analytical techniques. The result in Eq.(\ref{comparNaka}) is
known to be exact in the limits $\rho_s \ll 1$ and $\rho_s \sim 1$, and serves as a very good approximation for the
self-diffusion coefficient in hard-core lattice gases of arbitrary density \cite{kehr}.

\section{Conclusion}

To conclude,  we have studied analytically the
 intrinsic frictional properties
 of 2D adsorbed monolayers,
composed of mobile hard-core particles  
undergoing continuous exchanged with the vapor. 
Our analytical approach has been based on
the master equation, describing the time
 evolution of the system, 
which has allowed us to evaluate a system of coupled
dynamical equations 
for the tracer particle
velocity and a 
hierarchy of correlation functions. 
To solve these coupled equations, we have invoked an approximate closure scheme
based on the decomposition of the
third-order correlation functions into a product of pairwise correlations, which has
been  
first 
introduced in Ref.\cite{burlatsky} for  a related
 model of a driven tracer particle dynamics in a one-dimensional lattice gas 
with conserved particles number. 
Within the framework of this approximation,
we have derived a system of coupled, discrete-space equations describing evolution 
of the density profiles in the adsorbed monolayer, as seen from the  moving
tracer, and its velocity $V_{tr}$.  We have shown that  the density profile  around the tracer is strongly
inhomogeneous: the local density of the adsorbed
 particles in front of the tracer is higher than the 
average and approaches the average value as an exponential
 function of the distance from the tracer. 
On the other hand, past the tracer 
the local density is always lower than the average, and depending on
whether the number of particles 
is explicitly conserved or not, the local density past the tracer
 may tend to the average value either as an exponential or even as an
 $\it algebraic$ function of the distance. The latter reveals 
especially strong memory effects and strong 
correlations between the particle distribution in the
environment and the carrier position. 
Next, we have derived a general force-velocity relation, which  defines the terminal velocity of the tracer particle
for arbitrary applied fields and arbitrary values of other system parameters. 
We have demonstrated next that in the limit of a vanishingly small external bias this relation attains a simple, but physically
meaningful form of the Stokes formula, which signifies that in this limit the frictional force exerted on the tracer  by
the adsorbed monolayer particles is viscous. Corresponding friction coefficient has been also explicitly determined. 
In addition, we estimated the self-diffusion
coefficient of the tracer in the absence of the field and showed that it reduces to the well-know result of
Refs.\cite{nakazato} and \cite{tahir} in the limit when the particles number is conserved.


\vskip 1cm

{\Large  \bf Figure Captions.}

\vskip 1cm

Fig.1. Adsorbed monolayer in contact with a vapor. 
Grey spheres denote the
monolayer (vapor) particles; the grey sphere with an arrow 
stands for the driven tracer
particle.

Fig.2. Singular points $z_i$  of the generating function $N(z)$ and the
integration contour (thick line).

Fig.3. Piecewise contour  ${\cal H}(n)$ of
the Hankel type.

\newpage

{\Large  \bf Appendix A}

\vskip 1cm

In this Appendix we present the details of calculation of different contributions to  the time evolution of the pair correlation
function 
 $k({\boldsymbol \lambda};t)$ (cf. Eq.(\ref{evolk})), associated with four lines in the master equation  (\ref{eqmaitresse}).
 
Consider first the contribution associated with the hopping motion of the adsorbed particles: 
\begin{eqnarray}
C_1({\boldsymbol \lambda})&\equiv&\sum_{\bf R_{tr},\eta}\sum_{\mu =1,2}\;\sum_{{\bf r}\neq{\bf R_{tr}}-{\bf e}_{\boldsymbol \mu},{\bf R_{tr}}}\eta({\bf R_{tr}}+{\boldsymbol \lambda})
  \left\{ P({\bf R_{tr}},\eta^{{\bf r},\mu};t)-P({\bf R_{tr}},\eta;t)\right\}\nonumber\\
&=&\sum_{\bf R_{tr},\eta}\sum_{\mu =1,2}\;\sum_{{\bf r}\neq{\bf R_{tr}}-{\bf e}_{\boldsymbol \mu},{\bf R_{tr}}}
\left\{ \eta^{{\bf r},\mu}({\bf R_{tr}}+{\boldsymbol \lambda})-\eta({\bf R_{tr}}+{\boldsymbol \lambda})\right\}P({\bf R_{tr}},\eta;t).\nonumber\\    
\end{eqnarray}
Now, one has then to distinguish between two possible situations:
\begin{enumerate}
\item [(a)] when  ${\boldsymbol \lambda}\neq{\bf e}_{\boldsymbol \nu}$, one has
\begin{eqnarray}
&&\sum_{\mu =1,2}\sum_{{\bf r}\neq{\bf R_{tr}}-{\bf e}_{\boldsymbol \mu},{\bf R_{tr}}}\left\{ \eta^{{\bf r},\mu}({\bf R_{tr}}+{\boldsymbol \lambda})-\eta({\bf R_{tr}}+{\boldsymbol \lambda})\right\}P({\bf R_{tr}},\eta;t)\nonumber\\
&=&\sum_{\mu =1,2}(\nabla_\mu+\nabla_{-\mu})\eta({\bf R_{tr}}+{\boldsymbol \lambda})\nonumber\\
&=&\sum_\mu\nabla_\mu\eta({\bf R_{tr}}+{\boldsymbol \lambda}),
\end{eqnarray}which yields, eventually, 
\begin{eqnarray}
C_1({\boldsymbol \lambda})=\sum_\mu\nabla_\mu k({\boldsymbol \lambda};t).
\end{eqnarray}
\item [(b)] when ${\boldsymbol \lambda}={\bf e}_{\boldsymbol \nu}$, i.e. at the sites adjacent to the tracer, we find
\begin{eqnarray}
&&\sum_{\mu =1,2} \sum_{{\bf r}\neq{\bf R_{tr}}-{\bf e}_{\boldsymbol \mu},{\bf R_{tr}}}\left\{ \eta^{{\bf r},\mu}({\bf R_{tr}}+{\bf e}_{\boldsymbol \nu})-\eta({\bf R_{tr}}+{\bf e}_{\boldsymbol \nu})\right\}P({\bf R_{tr}},\eta;t)\nonumber\\
&=&(\sum_\mu\nabla_\mu-\nabla_{-\nu})\eta({\bf R_{tr}}+{\bf e}_{\boldsymbol \nu}),
\end{eqnarray} 
which implies that
\begin{equation}
C_1({\bf e}_{\boldsymbol \nu})=(\sum_\mu\nabla_\mu-\nabla_{-\nu})k({\bf e}_{\boldsymbol \nu};t).
\end{equation}
\end{enumerate}
Finally, using the  Kroneker-delta $\delta_{{\boldsymbol \lambda},{\bf
e}_{\boldsymbol \mu}}$, we can generalize both results for arbitrary ${\boldsymbol
\lambda}$:
\begin{equation}
C_1({\boldsymbol \lambda})=(\sum_\mu\nabla_\mu-\delta_{{\boldsymbol \lambda},{\bf e}_{\boldsymbol \mu}}\nabla_{-\mu})k({\boldsymbol \lambda};t).
\end{equation}

Next, we turn to the contibution stemming out of random biased hopping motion
of the tracer particle. This reads
\begin{eqnarray}
C_2({\boldsymbol \lambda})&\equiv&\frac{4 \tau^*}{\tau}\sum_{{\bf R_{tr}},\eta} \sum_{\mu} p_\mu\eta({\bf R}_{tr}+{\bf e}_{\boldsymbol \nu})\{\left(1-\eta({\bf R_{tr}})\right)P({\bf R_{tr}}-{\bf e}_{\boldsymbol \mu},\eta;t)\nonumber\\
&&\;\;\;\;\;\;\;\;\;\;\;
-\left(1-\eta({\bf R_{tr}}+{\bf e}_{\boldsymbol \mu})\right)P({\bf R_{tr}},\eta;t)\}\nonumber\\
&=&\frac{4 \tau^*}{\tau}\sum_{{\bf R_{tr}},\eta} \sum_{\mu} p_\mu(1-\eta({\bf R_{tr}}+{\bf e}_{\mu}))\left\{\eta({\bf R}_{tr}+{\bf e}_{\boldsymbol \nu}+{\bf e}_{\boldsymbol \mu})-\eta({\bf R}_{tr}+{\bf e}_{\boldsymbol \nu})\right\}P({\bf R_{tr}},\eta;t)\nonumber\\
&=&\frac{4 \tau^*}{\tau}\sum_{{\bf R_{tr}},\eta} \sum_{\mu} p_\mu\left(
1-\eta({\bf R_{tr}+e}_{\boldsymbol \mu})\right)\nabla_\mu\eta({\bf R_{tr}}+{\boldsymbol \lambda})P({\bf R_{tr}},\eta;t),
\end{eqnarray}

Further on, we consider the contribution associated with desorption
 of the monolayer particles:
\begin{eqnarray}
C_3({\boldsymbol \lambda})&\equiv&4 g\sum_{\bf R_{tr}}\sum_{{\bf r}\neq{\bf R_{tr}}}\sum_\eta\eta({\bf R_{tr}}+{\boldsymbol \lambda})\left\{\left(1-\eta({\bf r})\right)P({\bf R_{tr}},\hat{\eta}^{{\bf r}};t)-\eta({\bf r})P({\bf R_{tr}},\eta;t)\right\}\nonumber\\
&=&4 g\sum_{\bf R_{tr}}\sum_{{\bf r}\neq{\bf R_{tr}}}\sum_\eta\eta({\bf r})\{\hat{\eta}^{\bf r}({\bf R_{tr}}+{\boldsymbol \lambda})-\eta({\bf R_{tr}}+{\boldsymbol \lambda})\}P({\bf R_{tr}},\eta;t)\nonumber\\
&=&4 g\sum_{\bf R_{tr}}\sum_\eta\eta({\bf R_{tr}}+{\boldsymbol \lambda})(1-2\eta({\bf R_{tr}}+{\boldsymbol \lambda}))P({\bf R_{tr}},\eta;t)
\end{eqnarray}
Taking into account that $\eta({\bf R})$ assumes only two values - $0$ and $1$, and hence, that $\eta^2({\bf R}) = \eta({\bf R})$, we have
\begin{equation}
C_3({\boldsymbol \lambda})
=-4 gk({\boldsymbol \lambda};t).
\end{equation}
Lastly, the contribution due to adsorption of the particles from the vapor phase onto the
lattice  reads
\begin{eqnarray}
C_4({\boldsymbol \lambda})&\equiv& 4 f\sum_{\bf R_{tr},\eta}\sum_{{\bf r}\neq{\bf R_{tr}}} \eta({\bf R_{tr}}+{\boldsymbol \lambda})\left\{\eta({\bf r})P({\bf R_{tr}},\hat{\eta}^{{\bf r}};t)-(1-\eta({\bf r}))P({\bf R_{tr}},\eta;t)\right\}\nonumber\\
&=&4f\sum_{\bf R_{tr},\eta}\sum_{{\bf r}\neq{\bf R_{tr}}}(1-\eta({\bf r}))\{\hat{\eta}^{\bf r}({\bf R_{tr}}+{\boldsymbol \lambda})-\eta({\bf R_{tr}}+{\boldsymbol \lambda})\}P({\bf R_{tr}},\eta;t)\nonumber\\
&=&4f\sum_{\bf R_{tr},\eta}(1-\eta({\bf R_{tr}}+{\boldsymbol \lambda}))(1-2\eta({\bf R_{tr}}+{\boldsymbol \lambda}))P({\bf R_{tr}},\eta;t)
\end{eqnarray}
Again, since $\eta^2({\bf R}) = \eta({\bf R})$, the latter equation reduces to
\begin{equation}
C_4({\boldsymbol \lambda})=-4fk({\boldsymbol \lambda};t)+4f.
\end{equation}
Summing up all four contributions, we arrive at the evolution equation in Eq.(\ref{evolk}).

\newpage

{\Large  \bf Appendix B}

\vskip 1cm

We present here a derivation of the asymptotic behavior of the
density profiles at large separations from the stationary moving tracer. 
The generating function $N(z)$, defined by Eq.(\ref{expressionN}), has
 the roots  $z_i$ satisfy Eqs.(\ref{2ddefz1}) to (\ref{2ddefz4}).
Introducing the cuts depicted on Fig.2, one observes that $N(z)$
is an analytic function in the annular region  of  inner
radius  $z_2$ and the outer radius  $z_3$. In consequence, density deviations $h_{n,0}$ from the equilibrium value $\rho_s$
at positions ${\boldsymbol \lambda} = {\boldsymbol e}_1 n$
are given by the Cauchy formula
\begin{equation}
h_{n,0}=\frac{1}{2i\pi}\oint\frac{N(z)}{z^{n+1}}{\rm d}z,
\label{intcontour}
\end{equation} 
where the  contour of the integration is a positively oriented
circle, centered around O, of radius $R$ such that  $z_2 < R < z_3$ (see Fig.2).

\paragraph{I. Asymptotic behavior of $h_{n,0}$ in the limit $n\to + \infty$.}

Following closely the reasonings of Flajolet et al. \cite{darboux1}, we notice 
that the asymptotical behavior of
$h_{n,0}$ in the limit $n\to+\infty$ is supported by the behavior of the generating function $N(z)$ in the vicinity of $z = z_i$, where
$z_i$ is the first root encountered when
one tries to deforme the  contour in the integral (\ref{intcontour})
by increasing its radius. In our case, the relevant root is $z_3$. Now, we have to
choose a contour of integration that comes close enough to ``capture'' the
behavior of $N(z)$ in the vicinity of this leading singularity. Let us
formalize this idea.

We begin by expanding  $N(z)$ in the vicinity of $z_3$, which gives   
\begin{equation}
N(z)=C(z_3-z)^{-1/2}-h({\bf e_2})+{\mathcal O}(\sqrt{z_3-z}),
\end{equation}
where $C$ is a constant defined by
\begin{eqnarray}
C&\equiv&\frac{z_3(z_3-1)\Big(A_1h({\bf e_1})+\rho_s(A_1-A_{-1})\Big)+(1-z_3)\Big(A_{-1}h({\bf e_{-1}})-\rho_s(A_1-A_{-1})\Big)}{\sqrt{8A_2z_3}}\nonumber\\
&\times&\left\{\left(\frac{\alpha/2-A_2}{A_{-1}}\right)^2-\frac{A_1}{A_{-1}}\right\}^{-1/4}.
\end{eqnarray}
Then, Eq.(\ref{intcontour}) attains the form
\begin{equation}
\label{su}
h_{n,0}=\frac{C}{2i\pi}\oint\frac{(z_3-z)^{-1/2}}{z^{n+1}}{\rm d}z-\underbrace{\frac{h({\bf e_2})}{2i\pi}\oint\frac{1}{z^{n+1}}{\rm d}z}_{=0}+\frac{1}{2i\pi}\oint\frac{{\mathcal O}(\sqrt{z_3-z})}{z^{n+1}}{\rm d}z.
\end{equation}
Consider next the first term on the rhs of Eq.(\ref{su}), i.e.,  
\begin{equation}
D_n\equiv\frac{C}{2i\pi}\oint\frac{(z_3-z)^{-1/2}}{z^{n+1}}{\rm d}z.
\label{contint}
\end{equation}
One notices first that the contour of integration in Eq.(\ref{contint})
may be replaced by the piecewise contour  ${\cal H}(n)$ of
the Hankel type
(cf. Fig.3):
\begin{equation}
{\cal H}(n)={\cal H}^-(n)+{\cal H}^+(n)+{\cal H}^0(n),
\end{equation}  
where 
\begin{equation}
\begin{cases}
{\cal H}^-(n)=\{z=w-\frac{i}{n},\;w\geq1\}& \\
{\cal H}^+(n)=\{z=w+\frac{i}{n},\;w\geq1\}& \\
{\cal H}^0(n)=\{z=1-\frac{e^{i\phi}}{n},\;\phi\in[-\frac{\pi}{2},\frac{\pi}{2}\}&.
\end{cases}
\end{equation}
Changing next the variable of integration in Eq.(\ref{su}), $z=z_3(1+t/n)$, we have that $D_n$ obeys
\begin{equation}
D_n=n^{-1/2}z_3^{-n}\frac{Cz_3^{-1/2}}{2i\pi}\int_{{\cal H}}(-t)^{-1/2}\left(1+\frac{t}{n}\right)^{-n-1}{\rm d}t
\label{integrand2d}
\end{equation}
Further on, expanding the kernel
\begin{equation}
\left(1+\frac{t}{n}\right)^{-n-1}=e^{-(n+1)\ln(1+t/n)}=e^{-t}\Big(1+\frac{t^2-2t}{2n}+\ldots\Big),
\end{equation}
we observe that the integrand in  Eq.(\ref{integrand2d}) converges to
$(-t)^{-1/2}e^{-t}$, which is just the kernel appearing in
the Hankel's representation of the Gamma function. Turning to the limit $n \to + \infty$, and using an approximation  
\begin{equation}
\left(1+\frac{t}{n}\right)^{-n-1}=e^{-t}\left(1+{\mathcal O}\left(\frac{1}{n}\right)\right),
\end{equation}
in the integral in Eq.(\ref{integrand2d}),  we find eventually that 
\begin{eqnarray}
D_n&=&\frac{C}{\sqrt{z_3}\Gamma(1/2)}\frac{1}{\sqrt{n}z_3^n}\left(1+{\mathcal O}\left(\frac{1}{n}\right)\right)\nonumber\\
&=&\frac{C}{\sqrt{z_3\pi 
}}\frac{1}{\sqrt{n}z_3^n}\left(1+{\mathcal O}\left(\frac{1}{n}\right)\right)
\end{eqnarray}

Next, using essentially the same kind of arguments, it is easy now to show that
\begin{equation}
\frac{1}{2i\pi}\oint\frac{{\mathcal O}(\sqrt{z-z_3})}{z^{n+1}}{\rm d}z={\mathcal O}\left(\frac{1}{n^{3/2}}\right),
\end{equation}
which implies that in the limit 
$n\to+\infty$, the leading behavior is given by
\begin{equation}
h_{n,0}\sim\frac{C}{\sqrt{z_3\pi 
}}\frac{1}{\sqrt{n}z_3^n},
\end{equation}
which is exactly the result in Eqs.(\ref{2ddevh}) to (\ref{2dlongueurdevant}).

\paragraph{II. Asymptotical behavior of  $h_{n,0}$ in the limit $n\to-\infty$.}

To study asymptotical behavior of the density profiles at large separations past stationary moving tracer,
we proceed exactly in the same way as in the previous paragraph. First, we introduce the
generating function of the form
\begin{equation}
M(z)=\sum_{n=-\infty}^{+\infty}h_{-n,0}z^n
\end{equation} 
Since $M(z)$ obeys $ M(z)\equiv N(1/z)$, one immediately has that $M(z)$ is given explicitly by
\begin{eqnarray}
M(z)&=&\frac{(1-z)\Big(A_1h({\bf e_1})+\rho_s(A_1-A_{-1})\Big)+z(z-1)\Big(A_{-1}h({\bf
e_{-1}})-\rho_s(A_1-A_{-1})\Big)}{A_1\sqrt{(z-z_1^{-1})(z-z_2^{-1})(z-z_3^{-1})(z-z_4^{-1})}}\nonumber\\
&+&h({\bf e_2})\left(\sqrt{\frac{(z-z_2^{-1})(z-z_3^{-1})}{(z-z_1^{-1})(z-z_4^{-1})}}-1\right).
\end{eqnarray}
Hence, $M(z)$ is an analytic function in the annular region centered around O, of inner radius
$z_3^{-1}$ and the outer radius $z_2^{-1}$. 

Now,  the leading
singularity for $M(z)$ is in the vicinity of  $z=z_2^{-1}$. As explained in the text,
the nature of this singularity  $z_2^{-1}$ is different depending on whether the number of
particles in the monolayer is explicitly conserved or not.

\subparagraph{ A. Non conserved particles number.}

In this case $z_2^{-1}\neq1$ and hence, in the vicinity of  $z=z_2^{-1}$, the generating function $M(z)$ behaves as
\begin{equation}
\label{mi}
M(z)=E(z_2^{-1}-z)^{-1/2}-h({\bf e_2})+{\mathcal O}(\sqrt{z_2^{-1}-z}),
\end{equation} 
where $E$ is a constant, defined by
\begin{eqnarray}
E&\equiv&\sqrt{z_2}\frac{(1-z_2^{-1})\Big(A_1h({\bf e_1})+\rho_s(A_1-A_{-1})\Big)+z_2^{-1}(z_2^{-1}-1)\Big(A_{-1}h({\bf
e_{-1}})-\rho_s(A_1-A_{-1})\Big)}{\sqrt{8A_2}}\nonumber\\
&\times&\left\{\left(\frac{\alpha/2-A_2}{A_{1}}\right)^2-\frac{A_{-1}}{A_{1}}\right\}^{-1/4}
\end{eqnarray}
Inverting Eq.(\ref{mi}), we find then
the following asymptotical result:
\begin{equation}
h_{-n,0}\sim\frac{E\sqrt{z_2}}{\sqrt{\pi}}\frac{1}{\sqrt{|n|}}\frac{1}{z_2^n},
\end{equation}
which  corresponds to our Eqs.(\ref{2dderh}) to (\ref{2dlongueurderriere}).

\subparagraph{ B. Conserved particles number.}

In this case, $z_2^{-1}=1$ and consequently, one has that $M(z)$ admits the following expansion 
in the vicinity of  $z=z_2^{-1}$,
\begin{equation}
\label{m}
M(z)=-h({\bf e_2})+G(1-z)^{1/2}+{\mathcal O}\Big((1-z)^{3/2}\Big),
\end{equation}
where $G$ is the constant given by
\begin{equation}
G\equiv\frac{1}{2}\left(\frac{\Big(A_1h({\bf e_1})+\rho_s(A_1-A_{-1})\Big)-\Big(A_{-1}h({\bf
e_{-1}})-\rho_s(A_1-A_{-1})\Big)}{\sqrt{A_2(A_1-A_{-1})}}+h({\bf e_2})\sqrt{A_1-A_{-1}}{A_2}\right)
\end{equation}
Inverting Eq.(\ref{m}), we find that in the conserved particles number case, the asymptotical behavior of the density profile at
large separation past the stationary moving tracer particle obeys
\begin{equation}
h_{-n,0}=-\frac{G}{\pi n^3}\left(\frac{1}{2}+\frac{3}{16n}+{\mathcal O}\left(\frac{1}{n^2}\right)\right),
\end{equation}
i.e. the behavior described by our  Eqs.(\ref{algebrique}) and (\ref{2dconservampliderriere}).

\end{document}